\documentstyle[12pt]{article}
\normalbaselines
\begin{document}

\title{
 Symmetry properties of massive gauge  theories
 in nonlinear background gauges:
  Background dependence of Green functions
     }
     
\author{ B. Geyer \\ 
    Naturwissenschaftlich-Theoretisches Zentrum and\\
     Institut f\"ur Theoretische Physik, Universit\"at Leipzig,\\
          Augustusplatz 10-11, D-04109  Leipzig, Germany\\
     and \\     
   D. M\"ulsch\\
     Wissenschaftszentrum Leipzig e.V. \\
        Goldschmidtstr. 26, D-04109 Leipzig, Germany
     }
\maketitle

\begin{abstract}
Nonabelian gauge theories with a generic background field
$A_\mu$ in nonlinear gauges due to Delbourgo and Jarvis are
investigated.
The $A_\mu$-dependence is completely determined by the help
 of a linear differential equation which obtaines 
from the Kluberg-Stern-Zuber 
and the Lee identity.
Its integration leads to a relation between the 
one-particle irreducible vertex functional in the background
field $A_\mu$ 
and the corresponding functional for $A_\mu = 0$. An 
analogous relation holds for the generating functional of the
complete Green functions which, after restriction to 
physical Green functions, is used to confirm a result obtained
 by Rouet in the case of linear background gauge. 
\end{abstract}

\section{Introduction}

Quantum field theories with classical background configurations 
receive
growing attention, either because the full quantized theory is
far from being 
well defined as is the case for quantum gravity or, because of
the nontrivial 
topological structure of the classical theory, e.g. for
chromodynamics where instanton solutions give rise to a
complicated
structure of the QCD vacuum. This raises the question how the
renormalization
properties of the theory with background field are possibly  changed and
whether their Green
functions  may  be related to the Green functions without them. 

Here, this problem
will be considered for nonabelian gauge theories with generic
background field
$A_\mu$ within the algebraic BRST approach \cite{BRST} where
renormalizability of the theory
has to be proven by showing that the symmetries of the theory at
quantum level
are free of anomalies.
At first, this approach has been used by Kluberg-Stern and Zuber 
\cite{KSZ}. They also pointed out that the classical action 
\begin{eqnarray}
\label{YM}
 S_{\rm YM}(A) = - (2 g)^{-2} 
\int d^4x\, {\rm Tr}\bigr(F_{\mu\nu}(A + Q) F^{\mu\nu}(A + Q)
\bigr),
\end{eqnarray}
with field strenght $F_{\mu\nu}(A + Q) = 
\partial_\mu (A_\nu + Q_\nu) - \partial_\nu (A_\mu + Q_\mu) + 
[A_\mu + Q_\mu, A_\nu + Q_\nu]$,
is invariant under 
 two different kinds of transformations called type I and type
II:
\begin{eqnarray}
\label{I1}
\hbox{type I}:
\qquad
\delta A_\mu = D_\mu(A) \delta {\mit\Omega},
\qquad
\delta Q_\mu = [ Q_\mu, \delta {\mit\Omega} ], \\
\label{II1}
\hbox{type II}:
\qquad
\delta A_\mu = 0, 
\qquad\quad
\delta Q_\mu = D_\mu(A + Q) \delta {\mit\Omega}.               
\end{eqnarray}
If a {\em linear} background gauge is chosen, also the gauge
fixed action is
invariant under type I transformations, whereas type II
transformations have to
be changed into the BRST transformations.  Their Ward identities
are
 the (local) Kluberg-Stern--Zuber(KSZ) 
and Slavnov--Taylor(ST) identity, respectively. 
Obviously, there exists another kind of transformations, called
here type III,
\begin{eqnarray}
\label{III}
\hbox{type III}:
\qquad
\delta Q_\mu = D_\mu(Q) \delta {\mit\Omega},
\qquad
\delta A_\mu = [ A_\mu, \delta {\mit\Omega} ],  
\end{eqnarray}
which however cannot be required to hold unchanged for the
renormalized quantum action (see however equ. 
(\ref{lee}) below).

With the aim to determine the $A_\mu$--dependence of the Green
functions  
Rouet \cite{Rouet} used this formalism. In order to circumvent
the {\em nonlinear} (functional)
differential equation emerging for the generating functional 
$Z(A|J)$ he restricted the consideration
to the physical Green functions and obtained the relation
\begin{eqnarray}
\label{Rouet}
 Z_{\rm phys}(A|J) = {\rm exp}\left( - i (\hbar  z_Q)^{-1} 
\int d^4x\, {\rm Tr}(A^\mu J_\mu) \right) Z_{\rm phys}(0|J).
\end{eqnarray}
This shows that the on--shell amputated Green functions in the 
background field $A_\mu$ are deduced from the physical Green
functions 
in the vacuum by a mere translation of the (renormalized)
quantum field 
$z_Q Q_\mu$. 

In principle, to avoid IR--singularities it is necessary to 
provide the theory with a Higgs mechanism. However, there 
exist another possibilities.
In a recent paper \cite{GM2} we have shown that, if massive
gauge fields  are considered, 
e.g. with mass $m(s)=(1-s)m$, as it is  
necessary at least intermediately in the framework of the 
BPHZL--renormalization \cite{BPHZL}
for allowing an IR-regularization,
then only {\em nonlinear} gauges are permissible 
\footnote{The well known fact that massive gauge theories,
if renormalizable, violate unitarity does not matter here
since after carrying out all necessary subtractions the limit 
$s \rightarrow 1$ is taken.}. 
Therefore,  the Curci-Ferrari model \cite{CF} in the (nonlinear)
Delbourgo-Jarvis gauge \cite{DJNO} -- being the nonsingular
analogue
of the Landau gauge -- has been considered.
This gauge has the additional advantage that the theory is
invariant under BRST as well as
 anti-BRST transformations ${\sf\bf s}$ and ${\bar{\sf\bf s}}$, 
and, in the massive case, necessarily also under the 
$sp(2)$--transformations generated by
the FP operator, the DJNO and the anti-DJNO operator
${\sf\bf q}, {\sf\bf d}$ and ${\bar{\sf\bf d}}$, respectively 
(for the definitions see Chapter 2).

Now, in this paper we extend these considerations to include a
generic background 
field $A_\mu$. Imposing the gauge condition on the generating
functional for the
1PI vertex functions $ \Gamma (A|Q)$ it can be shown that 
a further local identity, being called Lee identity, emerges
which is
related to the type III symmetry above. Together with the
KSZ identity it  leads to a {\em linear} differential 
equation for the $A_\mu$--dependence of $ \Gamma (A|Q)$. 
Its solution relates $ \Gamma (A|Q)$ and $ \Gamma (0|Q)$; 
in addition, it can be shown that $A_\mu$ does not require any
further
$z$--factors 
and, therefore, the renormalization of  $ \Gamma (A|Q)$
is determined by that of $ \Gamma (0|Q)$.
 By Legendre transformation
 this gives a {\em complete} solution of the problem originally
considered by Rouet and,
after restriction to the physical subspace, confirmes his
result.
Furthermore, the method used here may be extended to consider
topological nontrivial background fields such as instantons and
merons.

\section{Massive gauge theory in nonlinear background gauge}

In its most symmetric form the theory is defined by the
following (classical) action (for its  explicit form see
(\ref{sclass1})):
\begin{eqnarray}
\label{sclass}
\hspace{-0.3cm}
S(A|g,m;\xi,\rho,\sigma) = 
     -(2g)^{-2} \int d^4 x 
     {\rm Tr}( F_{\mu\nu}(A+{\cal Q})F^{\mu\nu}(A+{\cal
Q}))\hspace{3cm}\\
   +\; ({\sf\bf s}_m {\bar{\sf\bf s}}_m + m^2) \int d^4 x 
            {\rm Tr}\Big(\hbox{$\frac{1}{2}$}{\cal Q}_\mu{\cal
Q}^\mu 
          +    \xi  {\bar{\cal C}}{\cal C}
          -    G_\mu {\cal Q}^\mu
          +    U {\bar{\cal C}} - {\bar U}{\cal C} 
          +    \hbox{$1\over 2$}\sigma G_\mu G^\mu  +  
  \rho {\bar U} U\Big)\nonumber
\end{eqnarray}
with
\begin{eqnarray}
\label{fields}
{\cal Q}_\mu   = Q_\mu + \sigma G_\mu, \quad
{\cal C}  =  C + \rho U, \quad
{\bar{\cal C}}   =  {\bar C} + \rho{\bar U},\quad
{\cal B}  =  B + (\zeta - {\bar{\zeta}})\,\rho\, S, 
\end{eqnarray}
where $Q_\mu, C, {\bar C}$ and $B$ are the gauge, ghost,
anti\-ghost
and Nakanishi-Lautrup field, respectively, and $G_\mu, U,
\bar{U}$ and $S$ are associated sources having the same
quantum numbers as these fields, respectively; 
$\xi$ is the gauge parameter, $\zeta$ an
arbitrary real parameter (${\bar{\zeta}} = 1 - \zeta$) 
reflecting the freedom in defining $ B$ according to 
${\cal B}={\bar{\zeta}}{\sf\bf s}_m{\bar{\cal C}}-
\zeta{\bar{\sf\bf s}}_m{\cal C}$,
whereas $\sigma$ and $\rho$ are to be introduced since the
dynamical fields
and the associated sources mix under renormalization.

The mass--dependent (anti--) BRST operators ${\sf\bf s}_m$
(and ${\bar{\sf\bf s}}_m$) are defined through
their action on the fields and the sources according to:
\begin{eqnarray}
\label{BRST}
{\sf\bf s}_m A_\mu = 0,\quad
{\sf\bf s}_m {\cal Q}_\mu =  D_\mu(A+{\cal Q}){\cal C},\quad
{\sf\bf s}_m {\cal C} = - (1/2) \left\{ {\cal C},{\cal C}
\right\},\nonumber\\
{\sf\bf s}_m {\bar{\cal C}}=
     {\cal B}-\zeta\left\{ {\cal C},{\bar{\cal C}}
\right\},\quad
{\sf\bf s}_m {\cal B}=\zeta\left[ {\cal B}+
     {\bar{\zeta}}\left\{ {\cal C},{\bar{\cal C}} \right\},
                          {\cal C} \right]- m^2 {\cal C},
\\
\label{BRST1}
{\sf\bf s}_m G_\mu =P_\mu,\quad {\sf\bf s}_m P_\mu  = 0,\quad
{\sf\bf s}_m U        = R,\quad {\sf\bf s}_m R  = 0,\nonumber\\
{\sf\bf s}_m \bar{U}  = S,\quad {\sf\bf s}_m S  = - m^2 U,\quad
{\sf\bf s}_m \bar{P}_\mu = - m^2 G_\mu ,\quad
{\sf\bf s}_m \bar{R}  = - 2 m^2 \bar{U}. 
\nonumber
\end{eqnarray}
The corresponding relations defining the action of the anti-BRST
operator
${\bar{\sf\bf s}}_m$ are obtained by the following conjugation
${\sf\bf C}$ which leaves the classical action (\ref{sclass})
invariant
and, obviously, is an extension of the 
ghost--antighost conjugation:
\begin{eqnarray}
C \rightarrow \phantom{-}{\bar C},\quad 
U \rightarrow \phantom{-}{\bar U}, \quad
P_\mu \rightarrow \phantom{-}{\bar P}_\mu,\quad
R \rightarrow {\bar R}, \quad 
S \rightarrow -S, \quad 
\zeta \rightarrow {\bar{\zeta}},
\nonumber\\ 
{\bar C} \rightarrow -C, \quad 
{\bar U} \rightarrow -U, \quad 
{\bar P}_\mu \rightarrow -P_\mu,\quad\;
{\bar R} \rightarrow R, \quad 
(S \equiv {\bar S}), \quad
{\bar{\zeta}} \rightarrow \zeta; 
\end{eqnarray}
$Q_\mu, G_\mu$ and $B$ as well as $\xi$ are $\sf\bf C$--even.
The additional sources $\bar{P_\mu}$ and $\bar R$ together with
$S$
 and $\bar{U}$ couple to the
 nonlinear parts of the BRST transformations (\ref{BRST}) of the
fields
(see  (\ref{sclass1})) and
analogously for the $\sf\bf C$--conjugate fields and sources;
the source $G_\mu$ couples to 
${\sf\bf s}_m {\bar{\sf\bf s}}_m {\cal Q}_\mu$. 

For later convenience we introduce the
 Delbourgo-Jarvis-Nakanishi-Ojima\-(DJNO)
 transformations \cite{DJNO} defined by:
\begin{eqnarray}
\label{DJNO} 
{\sf\bf d}A_\mu =
{\sf\bf d}{\cal Q}_\mu  = {\sf\bf d} G_\mu = 0, \quad
{\sf\bf d}{\bar{\cal C}}= {\cal C}, \quad
{\sf\bf d} {\cal C}   = 0,\quad
{\sf\bf d}{\cal B}    = (\zeta - {\bar{\zeta}}){\cal C}^2,
\\
\label{DJNO1}
{\sf\bf d}{\bar{P_\mu}} = P_\mu, \quad {\sf\bf d} P_\mu =
0,\quad
{\sf\bf d}{\bar U}= U,\quad {\sf\bf d} U  = 0,
{\sf\bf d} S      = R,\quad {\sf\bf d} R  = 0,\quad
{\sf\bf d}{\bar R}= 2 S.
\end{eqnarray}
Again, the anti--DJNO transformations are obtain\-ed by 
$\sf\bf C$--conjugation.

This finishes the definition of our model
whose  action has the following explicit form:
\begin{eqnarray}
\label{sclass1}
\hspace{-0,8cm} 
& &S(A|g,m;\xi,\rho,\sigma) =\\
\hspace{-0,5cm}
&-& (2g)^{-2}\int d^4 x 
{\rm Tr}(F_{\mu\nu}(A+{\cal Q})F^{\mu\nu}(A+{\cal Q}))
\quad + \quad m^2\int d^4 x {\rm
Tr}\Big(\hbox{$\frac{1}{2}$}{\cal Q}_\mu{\cal Q}^\mu 
     +  \xi  {\bar{\cal C}}{\cal C}\Big)\nonumber\\
\hspace{-0,5cm}
&+& \int d^4 x {\rm Tr}\Big({\cal Q}^\mu D_\mu(A) {\cal B}
   + \zeta \left(D_\mu(A){\cal C}\right) D^\mu(A+{\cal Q}){\bar{\cal C}}
   - {\bar\zeta}\left(D_\mu(A){\bar{\cal C}}\right)D^\mu(A+{\cal Q}){\cal
C}\Big)\nonumber\\
\hspace{-0,5cm}
&+& \xi \int d^4 x {\rm Tr}\Big(m^2{\bar{\cal C}}{\cal C}
   + \hbox{$\frac{1}{2}$}\left({\cal B}
     - \zeta\left\{ {\cal C},{\bar{\cal C}} \right\}\right)^2
   + \hbox{$\frac{1}{2}$}\left({\cal B}
    + {\bar\zeta}\left\{{\bar{\cal C}}, {\cal
C}\right\}\right)^2\Big)\nonumber\\
\hspace{-0,5cm}
&+& \int d^4 x {\rm Tr}\Big(- G_\mu D^\mu(A+{\cal Q}){\cal B}
   + \left({\bar P}_\mu 
     - {\bar\zeta}\left[G_\mu, 
          {\bar{\cal C}}\right]\right)D^\mu(A+{\cal Q}){\cal C}-
\nonumber\\
& & \hspace{7cm}   - \left( P_\mu 
     - \zeta\left[G_\mu, 
     {\cal C}\right]\right)D^\mu(A+{\cal Q}){\bar{\cal C}}\Big)
\nonumber\\
\hspace{-0,5cm}
&+& \int d^4 x {\rm Tr}\Big(-S\left\{ {\cal C},{\bar{\cal C}}
\right\}
     + {\bar R}{\cal C}^2 + R{\bar {\cal C}}^2
   + {\bar U}\left[ {\cal B}+
     {\bar{\zeta}}\left\{{\cal C},{\bar{\cal C}}\right\}, {\cal
C}\right]
   -  U \left[ {\cal B}-
     {\zeta}\left\{{\cal C},{\bar{\cal C}}\right\}, {\bar{\cal
C}}\right]\Big)
\nonumber\\
\hspace{-0,5cm}
&+& \sigma \int d^4 x {\rm Tr}\Big(P_\mu{\bar P}^\mu
     -  \hbox{$\frac{1}{2}$}m^2 G_\mu G^\mu\Big)
\quad + \quad \rho \int d^4 x {\rm Tr}\Big(R{\bar R} - S^2\Big).
\nonumber
\end{eqnarray}

Let us now state the (classical) symmetries of this theory.
Again, as in the case $A_\mu= 0$, 
 the algebra generated by 
$\left\{{\sf\bf s}_m, {\bar{\sf\bf s}}_m;{\sf\bf d},
{\bar{\sf\bf d}},
 {\sf\bf q}\right\}$ 
is easily verified to be the superalgebra $osp(1,2)$ \cite{OSP}. 
It is given by the following $\sf\bf q$--graded commutation
relations
%
\begin{eqnarray}
\label{OSP}
{\sf\bf s}_m^2        = - m^2 {\sf\bf d},\quad\;
\left\{{\sf\bf s}_m, {\bar{\sf\bf s}}_m\right\} = m^2 {\sf\bf
q},\quad\;
\bar{\sf\bf{ s}}_m^2  = - m^2 {\bar{\sf\bf d}},\\
\left[{\sf\bf d}, {\sf\bf s}_m\right]      = 0,\quad\;
\left[{\bar{\sf\bf d}}, {\sf\bf s}_m\right] = - {\bar{\sf\bf
s}}_m,\quad\;
\left[{\sf\bf q}, {\sf\bf s}_m\right]      = \phantom{-}{\sf\bf
s}_m,\\
\left[{\bar{\sf\bf d}}, {\bar{\sf\bf s}}_m\right] = 0,\quad\;
\left[{\sf\bf d}, {\bar{\sf\bf s}}_m\right] = \phantom{-}{\sf\bf
s}_m,\quad\;
\left[{\sf\bf q}, {\bar{\sf\bf s}}_m \right]  = - {\bar{\sf\bf
s}}_m,\\
\label{OSP1}
\left[{\sf\bf q}, {\sf\bf d}\right]        =  2 {\sf\bf
d},\quad\;
\left[{\sf\bf q}, {\bar{\sf\bf d}}\right]  = -2 {\bar{\sf\bf
d}},\quad\;
\left[{\sf\bf d}, {\bar{\sf\bf d}}\right]  = -  {\sf\bf q}.
\end{eqnarray}

Using these relations it is easy to check that the action
$S(A|g,m;\xi,\rho,\sigma)$ is invariant under (anti--) BRST as
well 
as (anti--) DJNO transformations:
\begin{eqnarray}
\label{sym}
{\sf\bf s}_m S(A|g,m;\xi,\rho,\sigma)=0,\quad 
{\sf\bf d}\; S(A|g,m;\xi,\rho,\sigma)=0, \nonumber\\
{\bar{\sf\bf{ s}}}_m S(A|g,m;\xi,\rho,\sigma)=0,\quad 
{\bar{\sf\bf{ d}}}\; S(A|g,m;\xi,\rho,\sigma)=0;
\end{eqnarray}
invariance under $\sf\bf q$ is trivial since
$S(A|g,m;\xi,\rho,\sigma)$
has ghost number zero by definition. Furthermore, the (anti--)
DJNO
symmetry of the action according to (\ref{OSP}) 
 is a consequence of the (anti--) BRST invariance!

Obviously, in
the limit $m^2 = 0$ the usual algebra of nilpotent (anti--)
BRST operators 
${\sf\bf s}^2 = 0, \; \{{\sf\bf s},{\bar{\sf\bf{ s}}}\} = 0, \;
{\bar{\sf\bf{ s}}}^2 = 0$ appears, and the $sp(2)$--algebra 
(\ref{OSP1}), 
being now an independent requirement, decouples. Of course, 
a massless Yang-Mills theory in nonlinear gauge obtains
\footnote{As has been shown by Curci and Ferrari \cite{CF} in
the
massless case two independent gauge parameters $\xi, {\bar\xi}$
may
be introduced; but than (anti--) DJNO invariance cannot be
maintained.}.
However, it should be emphazised that the mass is not
really an independent parameter of the theory
 since the Slavnov-Taylor identities
 explicitly depend on $m$.

As in the case of  linear gauges,  the auxiliary field $B$
occurs quadratically
in the action,
but now through $(\zeta - {\bar\zeta}) B ({\bar C} C)$ it
couples to the
(anti--) ghost fields, which  have quartic selfcoupling;
 therefore $B$ does not decouple from the dynamics.
Finally we remark that, as is seen from (\ref{sclass1}),  the 
sources being coupled by $\sigma$ and
$\rho$ occure bilinear; despite looking strange this is
unavoidable for
a stable, $osp(1,2)$--invariant theory.

In addition to this $osp(1,2)$--symmetry, which is related to the
broken gauge, i.e. type II, symmetry , 
the action (\ref{sclass1}) is invariant (eventually modulo a mass term)
under (generalized) type
I and type III transformations defined as
\begin{eqnarray}
\label{typeI}
\hbox{type I}:
\quad
\delta A_\mu = D_\mu(A) \delta {\mit\Omega},
\;\;
\delta {\cal Q}_\mu = [ {\cal Q}_\mu, \delta {\mit\Omega} ], 
\;\;
\delta G_\mu = [ G_\mu, \delta {\mit\Omega} ], 
\;\;
\delta \Phi = [ \Phi, \delta {\mit\Omega} ],
\nonumber
\\ 
\label{typeIII}
\hbox{type III}:
\quad
\delta {\cal Q}_\mu = D_\mu({\cal Q}) \delta {\mit\Omega},
\;\;
\delta G_\mu = D_\mu(G) \delta {\mit\Omega}, 
\;\;
\delta A_\mu = [ A_\mu, \delta {\mit\Omega} ],
\;\;
\delta\Phi = [ \Phi, \delta {\mit\Omega} ],  
\nonumber
\end{eqnarray}
where $\Phi$ here and in the following 
denotes the remaining (dynamical or external)
fields not explicitly written down:
\begin{eqnarray}
\label{stypeI}
{\sf\bf\delta_I} S(A|g,m;\xi,\rho,\sigma)=0,\quad 
{\sf\bf\delta_{III}} S(A|g,m;\xi,\rho,\sigma)= m^2 \partial_\mu Q^\mu.
\end{eqnarray}
Whereas the first symmetry can be formulated as a Ward identity,
the KSZ identity
of the (renormalized) 
theory, the second one leads to the Lee--identity (\ref{lee}).

\section{Ward identities for the vertex functional $\Gamma$}

In order to construct the renormalized theory associated to the 
classical action (\ref{sclass}) the symmetries of that action
have to be 
expressed by corresponding Ward identities of the 1PI
functional. These identities hold, first of all, at lowest
perturbative
 order, i.e. for $\Gamma^0(A|g, m; \xi, \rho, \sigma)$;
they  are to be required to hold for the
renormalized 1PI functional $\Gamma (A|g, m; \xi, \rho,
\sigma)$.
Since these identities are purely algebraic requirements they
may be formulated for a generic functional, denoted also by
$\Gamma$, 
which depends on the same 
fields, sources and parameters:
\begin{eqnarray}
\label{WI1}
^A{\sf\bf K}\;\Gamma\; = 0 & &   {\rm (Kluberg-Stern-Zuber\;
identity)},\\
\label{WI2}
{\sf\bf S}_m (\Gamma) = 0  & &    {\rm (Slavnov-Taylor\;
identity)},\\
\label{WI3}
{\sf\bf D}\;(\Gamma) = 0   & &    {\rm (Delduc-Sorella\;
identity)},\\
\label{WI4}
{\sf\bf T}_\zeta(\Gamma) = 0 & & {\rm (Bonora-Pasti-Tonin\;
identity)},
\end{eqnarray}
and, in addition, the ${\sf\bf C}$-conjugated second and
third identity; the first  identiy and the last one
(fixing the definition of $\cal B$) are
 ${\sf\bf C}$--even. 

The above operations are defined in the following manner:
The Kluberg-Stern--Zuber (KSZ) operator extends the local
Ward operator 
${\sf\bf W} = \sum \left[\Phi,  \delta / \delta \Phi \right\}$ 
according to 
%
$^A{\sf\bf K}= D_\mu (A)  \delta / \delta A_\mu + {\sf\bf W}$
%
 with the commutator or anticommutator
depending on even or odd $\sf\bf q$--grading of $\Phi$,
respectively. 
Of course, identity (\ref{WI1}) is defined only for $A_\mu \neq
0$; after  integration
over spacetime we get the rigid Ward identity 
$\int d^4 x \left\{\left[A,  \delta / \delta A \right] + {\sf\bf
W}\right\}\Gamma = 0$.
The (bilinear) Slavnov-Taylor (ST) and Delduc-Sorella (DS)
\cite{DS} 
operations as usual are given by
%
${\sf\bf S}_m (\Gamma)= \int d^4x \sum {\rm Tr}
     \left(({\sf\bf s}_m \Phi)\delta\Gamma/\delta\Phi\right)
$
and
$
{\sf\bf D}(\Gamma)= \int d^4x \sum {\rm Tr}
     \left(({\sf\bf d}\Phi) \delta\Gamma/\delta\Phi
\right),
$
%
respectively, where the nonlinear parts of ${\sf\bf s}_m \Phi$
and 
${\sf\bf d} \Phi$ are to be replaced by the functional
derivative 
of $\Gamma$ with respect to the corresponding source 
${\bar P}_\mu, {\bar R}, S$ or $\bar U$ (e.g. 
$\zeta    \left[ {\cal B}+ {\bar{\zeta}}
     \left\{ {\cal C},{\bar{\cal C}} \right\}, {\cal C} \right]$
by $\zeta \delta\Gamma / \delta {\bar U}$).
The Bonora-Pasti-Tonin (BPT) operation \cite{BPT} is defined by
%
$
{\sf\bf T}_\zeta(\Gamma) =  \partial\Gamma/\partial\zeta
 - \int d^4x {\rm Tr} \left( \delta\Gamma/\delta S \cdot
                          \delta\Gamma/\delta B\right).
$

For linear gauges the action $S$, if appropriately chosen,
coincides with
the effective action in tree approximation
$\Gamma^0$. But in the case of nonlinear gauges this will not
be true. Namely, the ST operation applied to $S(A|g, m; \xi,
\rho,
\sigma)$  leads to the expression
$
{\sf\bf S}_m(S) = 2 \zeta{\bar{\zeta}}\rho \int d^4x {\rm Tr} 
   ( \delta S/ \delta B \cdot \delta S/ \delta {\bar C})
$
which vanishes only in the special case $ \zeta{\bar{\zeta}} =
0$.
Therefore, $S$  has to be changed in such a manner that this
unwanted term is cancelled. The explicit determination is very
cumbersome, it leads to the following essential result:
\begin{eqnarray}
\Gamma^0(A|g, m; \xi, \rho, \sigma) = S(A|g, m; \xi, \rho,
\sigma) 
    +\;  \frac{ 
\zeta\bar{\zeta}\rho}{1-4\zeta\bar{\zeta}\rho\xi}
          \int d^4x {\rm Tr}
     \left( \frac{\delta S}{\delta B}\frac{\delta S}{\delta
B}\right) .
\end{eqnarray}
As can be seen by an explicit computation
  this expression fulfills the
identities (\ref{WI1}) -- (\ref{WI4}) and, therefore,
 constitutes the correct 
starting point for perturbative renormalization. In addition, 
it holds 
$
\tau \delta \Gamma^0 / \delta B = \delta S / \delta B$ with 
$
 \tau = (1 -  \rho\xi).
$

Here, it should be remarked that $S(A|g, m; \xi, 0, 0)$ also 
fulfils the identities (\ref{WI1}) -- (\ref{WI4})
and, in fact, this restricted action
was the expression we formerly started from. But, as is easily
seen, it
is not stable against small perturbations, i.e. not all the
field monomials $\Delta$ introduced in such a way that
 $S + \epsilon\Delta$ 
 fulfills the identities (\ref{WI1}) -- (\ref{WI4}) 
up to order $O(\epsilon^2)$ can
be obtained from $S$ by a variation of the parameters of that
restricted action. Therefore it had to be extended by
introducing the $\rho$-- and $\sigma$--dependent combinations
(\ref{fields}) of the fields and the associated sources.

\section{Gauge and consistency conditions}

The auxiliary field $B$, despite of the fact that it does not 
decouple from the dynamics, is not a primary one. However, its
dynamics can be constrained to be the same as for the classical
action by the following gauge condition (which, in fact,  is the 
renormalized equation of motion):
\begin{eqnarray}
\label{gauge}
\tau\;\delta \Gamma/\delta B 
&=&
 \xi(2B + (\zeta - {\bar{\zeta}})\;\delta\Gamma / \delta S) +
     D_\mu(A)(G^\mu -{\cal Q} ^\mu)\\
& &+ \left[{\cal Q}_\mu, G^\mu \right] 
          + \left\{ U, {\bar {\cal C}} \right\} 
     - \left\{ {\bar U}, {\cal C} \right\}.\nonumber
\end{eqnarray}
Here, we emphazise that, contrary to the common belief,
(\ref{gauge})
is a well posed condition since its nonlinear content is
respected
by the functional derivation of $\Gamma$ with respect to the
source $S$.

Applying now $\delta/\delta B$ on the
ST identity an equation completely analogous to (\ref{gauge})
obtains
for $\delta\Gamma/\delta {\bar C}$ which is nothing else
than the antighost equation of motion:
\begin{eqnarray}
\label{antighost}
\hspace{-0.8cm}\tau\;\delta \Gamma/\delta{\bar C}
 = 
        \xi  (2m^2 C -  \delta\Gamma/\delta{\bar U})
     - D_\mu(A)\left((1-\sigma)P^\mu
          -\delta\Gamma/\delta{\bar P}_\mu\right)
     - \left[Q_\mu, P^\mu\right] \hspace{.8cm}\\
\hspace{.5cm}
     + \left[{\bar C}, R\right]
     + \left[{\bar U}, \delta\Gamma/\delta{\bar R} \right]
     + \left[G_\mu, \delta\Gamma/\delta{\bar P}_\mu \right]
      - \left[C, S  + \zeta  \delta\Gamma/\delta B    \right]
      + \left[U, B + \zeta \delta\Gamma/\delta S    \right];
\nonumber
\end{eqnarray}               
an analogous condition for the ghost equation of motion
obtains by $\sf\bf{ C}$--conjugation.

If now $\delta /\delta{\bar C}$ is applied on the anti--ST
identity we finally get a further independent condition, the
local Lee identity:
\begin{eqnarray}
\label{lee}
%
(D_\mu (A){\bf {\Delta}}^\mu(\sigma)  + {\sf\bf W})\Gamma
      = m^2 D_\mu (A) Q^\mu
\quad {\rm with}\quad
{\bf {\Delta}}^\mu(\sigma) \equiv
(1-\sigma)\frac{\delta}{\delta Q_\mu}
          +\frac{\delta}{\delta G_\mu}.
%
\end{eqnarray}

Surprisingly, this identity is independent of $\xi$, and also of 
$\zeta$ and $\rho$. It expresses a consistence condition
related to the Type III symmetry (which can be seen more
directly if the 
$A_\mu$--dependence of $\Gamma$ is taken into accont).
It should be remarked that in the case $A_\mu = 0$, where
the KSZ identity disappears, the integrated Lee identity
equals the rigid Ward identity.

Together with the identities (\ref{WI1}) -- (\ref{WI4}) the
conditions 
(\ref{gauge}) -- (\ref{lee})
constitute a basic set of algebraic relations which express the
symmetries to be fulfilled by the vertex functional. 
Additional consistency conditions cannot appear since
the linearized ST-- and DS--operations fulfil 
(anti--) commutation relations
analogous to (\ref{OSP}) -- (\ref{OSP1}); and the BPT-identity
is  not
independent, but follows by the help of
$\partial {\sf\bf S}_m (\Gamma) / \partial\zeta = 0$.

As a consequence of the gauge and consistency conditions
(\ref{gauge}) -- (\ref{lee}) the $z$--factors of the
theory are nonlinearly related (see \cite{GM2}); especially
it holds
$1 - \sigma z_\sigma = (1 - \sigma) z_Q$ which is relevant
for checking (\ref{Rouet}) using 
(\ref{bcZsol}) and (\ref{bcG}) below.

Up to now we were concerned with strictly massive theories. If 
massless fields were to be involved, subtractions at zero momenta would
generate spurious infrared divergences due to the occurence of 
propagators with vanishing mass. However, the subtraction procedure 
 of Lowenstein and Zimmermann \cite{BPHZL} 
is free of spurious infrared  divergences. In this procedure a mass
$m^2 (s - 1)^2$ is put in the free propagator where $s$ 
($0 \leq s \leq 1$) plays the role of an additional infrared subtraction
parameter. Lowenstein, Zimmermann and Speer \cite{LZS} have proved that,
for non-exceptional external momenta, the limit $s \rightarrow 1$ defines
a massless theory. This is estabished by the Lowenstein-Zimmermann
equation 
 %
$m \partial\Gamma / \partial m  = O(s - 1)$.
%
 %
The differential operator $m\partial/\partial m$ applied to 
$\Gamma$ generates,
up to terms vanishing for $s = 1$, symmetrical insertions. 
Therefore, expanding
$m\partial/\partial m$ on a basis of symmetrical insertions and choosing
suitable  $s$-independent renormalization conditions  it can be proven
that all coefficients in this expansion vanish at any order of perturbation
theory (thereby it is supposed that the coefficients vanish at 
zeroth order). Thus for $s = 1$ one gets a functional $\Gamma$ which  
no longer  depends on the auxiliary mass $m$.

\section{Background dependence of Greens functions}

Now, combining the KSZ-- and the Lee--identity we obtain a 
functional differential equation for $\Gamma$ 
where the derivative w.r. to
$A_\mu$ is {\em linear} and related to the derivatives w.r. to $Q_\mu$
and $G_\mu$, namely
\begin{eqnarray}
\label{bc}
\left(\frac{ \delta}{\delta A_\mu} 
     - (1 - \sigma)\frac{ \delta}{\delta Q_\mu}   
     -\frac{ \delta}{\delta G_\mu} \right) \Gamma(A|Q,G,\Phi) 
     + m^2 Q^\mu = 0.
\end{eqnarray}
More precisely, originally this equation (\ref{bc}) occurs 
 with the background covariant derivative $D_\mu(A)$ acting from 
the left. 
Therefore, its solution is determined up to a background
covariant
constant functional, like $D_\nu(A) F^{\mu\nu}(A)$.
However, since the classical action (\ref{sclass1}) being
the tree
approximation of  $\Gamma$ does not depend on such terms, also
the
solution of  (\ref{bc}) should not depend on them.

With this boundary condition the solution of  (\ref{bc})  is
given by
\begin{eqnarray}
\label{bcsol}
 \Gamma(A|Q,G,\Phi) = 
     \exp (A\cdot {\bf\Delta}) \Gamma(0|Q,G,\Phi)
     - m^2 \{1 + \hbox{$\frac{1}{2}$}(A\cdot {\bf\Delta})\}
     (A\cdot Q),
\end{eqnarray}
with the obvious abbreviation
$
(X\cdot Y) = \int d^4x {\rm Tr}(X_\mu Y^\mu).
$
From this it follows that -- up to $m^2$--dependent term  --
the  $A_\mu$--dependence of $ \Gamma(A|Q,G,\Phi)$ is given by a
shift of $Q$ and $G$ in  $\Gamma(0|Q,G,\Phi)$, namely
\begin{eqnarray}
\label{Gamma}
 \Gamma(A|Q,G,\Phi) =  \Gamma(0|Q + (1 - \sigma )A,G + A,\Phi) 
     - m^2 \left( A\cdot (Q + 
     \hbox{$\frac{1}{2}$}(1 - \sigma)A)\right).
\end{eqnarray}

By this approach we obtained a complete description of the
background
field dependence for the vertex functional. It should be noted
that this
was possible because (\ref{bc}) is a linear differential
equation, and this 
results from the fact that a nonlinear gauge has been chosen.
Let us further remark that our derivation is purely algebraic. 
Therefore, the solution holds to any order of perturbation
theory. Furthermore,
without much effort it can be shown that $A_\mu$ will not be
renormalized.
Therefore, it is sufficient to prove renormalizability of 
$\Gamma (0|Q,G,\Phi)$.

The obtained result may be transfered to the 
generating functional $Z$ of the complete Green functions by
applying a Legendre transformation. Therefore let us
introduce
 %
%
$
W(A|J,G, \Phi) = i \hbar \ln Z(A|J,G,\Phi),
$
%
 %
being defined through
%
%
$
\Gamma (A|Q,G, \Phi) + W(A|J,G, \Phi) + (J\cdot Q) + (K\cdot B)
 + (L\cdot {\bar C}) - ({\bar L}\cdot C) = 0.
$
%
 %
Then the differential equation corresponding to (\ref{bc}) is
given by
\begin{eqnarray}
\label{bcZ}
\left(\frac{ \delta}{\delta A_\mu} -
     \frac{ \delta}{\delta G_\mu} + m^2 \frac{ \delta}{\delta J_\mu}
     - (1 - \sigma)\frac{J^\mu}{i\hbar}\right)Z(A|J,G,\Phi) = 0.
\end{eqnarray}

In the same manner as above the solution for the
Greens functions is obtained
\begin{eqnarray}
\label{bcZsol}
Z(A|J,G,\Phi) = \exp \{(1-\sigma)\left(A\cdot( J 
               - \hbox{$\frac{1}{2}$} m^2 A)\right)/i\hbar\}
                    Z(0|J - m^2 A,G+A,\Phi).
\end{eqnarray}
Furthermore, 
 comparing the Lee identity for $G_\mu\neq 0; \Phi = 0,$ 
\begin{eqnarray}
 \left(\frac{1-\sigma}{i\hbar}\partial_\mu J^\mu
     - m^2 \partial_\mu \frac{\delta}{\delta J_\mu}
     + D_\mu(G)  \frac{\delta}{\delta G_\mu}
     + [J_\mu, \frac{\delta}{\delta J_\mu}]\right)Z(0|J,G,0)
= 0,
\nonumber
\end{eqnarray}
with the corresponding one for $G_\mu = 0$ (being deduced
from the rigid Ward identity)
\begin{eqnarray}
\left((i\hbar z_Q)^{-1}\partial_\mu J^\mu
     + [J_\mu, \delta/\delta J_\mu]\right)Z(0|J,0,0) = 0,
\nonumber
\end{eqnarray}
the $G_\mu$--dependence 
of the renormalized functional $Z(0|J,G,0)$
may be determined explicitly:
\begin{eqnarray}
\label{bcG}
Z(0|J,G,0) = \exp \left\{ (\sigma z_\sigma/i\hbar z_Q)
\left((J  + 
     \hbox{$\frac{1}{2}$}m^2 G)\cdot G\right)\right\}
Z(0|J,0,0). 
\end{eqnarray}
Combining (\ref{bcZsol}) and (\ref{bcG}) we obtain
  for the physical Green functions in Landau gauge ($\xi
= 0$),
being restricted by the gauge condition 
$D_\mu(A) \delta Z_{\rm phys}(A|J)/\delta J_\mu = 0$ and
$\zeta$--independence  through
$\partial Z_{\rm phys}(A|J)/\partial \zeta = 0$, in the limit
 $m^2 =0, G_\mu =0$ exactly Rouet's result 
 equ. (\ref{Rouet}) above.

At this stage we have to comment that in paper \cite{Rouet}
the fact  has been overlooked that the verification of equ. (10) of
that paper presupposes an assumption which is not proved.
By our method, being completely different from the one used
by Rouet, this gap has been closed and his result is
now verified.

\section{Concluding remarks}
 
Here the Curci-Ferrari model of massive
gauge fields in nonlinear gauge has been generalized 
to the presence of a generic background field -- not
necessarily being a solution of the field equations.
The symmetries of the model are enlarged due to the
appearance of the background field. From the KSZ-- and the
Lee identity a linear differential equation for 1PI vertex
functional and, after Legendre transformation,
 an analogous one for the generating functional
of the Greens functions is obtained. Their solutions
relate the vertex resp. Green functions with background
field $A_\mu$ to the corresponding one for
$A_\mu = 0$. In particular, it is shown that the on--shell
amputated Green functions in the background field $A_\mu$
are deduced from the physical Green functions in the vacuum
by a mere translation of the gauge field; this verifies
a former result of Rouet.

Furthermore, it is claimed that $A_\mu$ is not renormalized
and, therefore, the problem of renormalization of the theory
with background is reduced to the corresponding problem  
of the theory without background field. This, however, has
been shown in an earlier paper \cite{GM1}. There, it is
proved, that only three independent $z$--factors appear,
namely $z_Q, z_B$ and $z_g$ (see also \cite{GM2}). A more
detailed presentation of the above results will be given 
elsewhere.

Finally let us remark that this method of enlarging the
symmetries can be applied also to the special cases of
instanton and meron configurations. Despite the fact that in
these cases additional zero modes appear which are to be
considered in the same spirit as has been done for the gauge zero modes
the formalism introduced here works with some additional technicalities.

\end{document}